\documentclass[10pt,conference]{IEEEtran}

\usepackage{booktabs}
\usepackage[utf8]{inputenc} 
\usepackage[T1]{fontenc} 
\usepackage{amsmath,amssymb} 
\usepackage{graphicx} 
\usepackage{cite} 
\usepackage{pdflscape}
\usepackage{subcaption}
\usepackage{balance} 
\usepackage{physics}
\usepackage{quantikz}
\usepackage{adjustbox}

\usepackage{hyperref,xcolor}
\definecolor{ionqorange}{HTML}{FF5000}
\hypersetup{
    colorlinks=true,
    linkcolor=ionqorange,    
    citecolor=ionqorange,    
    urlcolor=ionqorange      
}

\IEEEoverridecommandlockouts 

\title{A New Hybrid Quantum-Classical Algorithm for Solving the Unit Commitment Problem
\thanks{
This research used resources of the Oak Ridge Leadership Computing Facility, which is a DOE Office of Science User Facility supported under Contract DE-AC05-00OR22725. This work was supported by the U.S. Department of Energys' Grid Modernization Initiative (GMI) Grid Research, Integration, and Deployment for Quantum (GRID-Q) project. This manuscript has been authored by UT-Battelle, LLC under Contract No. DE-AC05-00OR22725  with the U.S. Department of Energy. The United States Government retains and the publisher, by accepting the article for publication, acknowledges that the United States Government retains a non-exclusive, paid-up, irrevocable, world-wide license to publish or reproduce the published form of this manuscript, or allow others to do so, for United States Government purposes. The Department of Energy will provide public access to these results of federally sponsored research in accordance with the DOE Public Access Plan (https://www.energy.gov/doe-public-access-plan).
}}

\author{
\IEEEauthorblockN{
        Willie Aboumrad\IEEEauthorrefmark{1},
        Phani R V Marthi\IEEEauthorrefmark{2},
        Suman Debnath\IEEEauthorrefmark{2},
        Martin Roetteler\IEEEauthorrefmark{1},
        Evgeny Epifanovsky\IEEEauthorrefmark{1}\medskip
    }
    \IEEEauthorblockA{\IEEEauthorrefmark{1}IonQ Inc., 4505 Campus Dr, College Park, MD 20740, USA}
    \IEEEauthorblockA{\IEEEauthorrefmark{2}Oak Ridge National Laboratory, 1 Bethel Valley Rd, Oak Ridge, TN 37830, USA}
}

\begin{document}

\maketitle

\begin{abstract}
Solving problems related to planning and operations of large-scale power systems is challenging on classical computers due to their inherent nature as mixed-integer and nonlinear problems. Quantum computing provides new avenues to approach these problems. We develop a hybrid quantum-classical algorithm for the Unit Commitment (UC) problem in power systems which aims at minimizing the total cost while optimally allocating generating units to meet the hourly demand of the power loads. The hybrid algorithm combines a variational quantum algorithm (VQA) with a classical Bender’s type heuristic. The resulting algorithm computes approximate solutions to UC in three stages: i) a collection of UC vectors capable meeting the power demand with lowest possible operating costs is generated based on VQA; ii) a classical sequential least squares programming (SLSQP) routine is leveraged to find the optimal power level corresponding to a predetermined number of candidate vectors; iii) in the last stage, the approximate solution of UC along with generating units power level combination is given. To demonstrate the effectiveness of the presented method, three different systems with 3 generating units, 10 generating units, and 26 generating units were tested for different time periods. In addition, convergence of the hybrid quantum-classical algorithm for select time periods is proven out on IonQ's Forte system. 
\end{abstract}


\section{Introduction}

The unit commitment (UC) problem aims to minimize the total cost of meeting specified hourly power loads by determining the optimal power level of each generating unit during each hour. UC is a complex power system problem characterized by exponential growth in computational complexity; as the number $M$ of time periods and the number $N$ of generators increases, the number of possible solutions grows like $(2^N-1)^M$ \cite{UC_1}. For example, in the case of the IEEE 9500-bus feeder \cite{UC_2} where $N = 15$, the total number of solutions in a $24$-hour period is $2.3 \cdot 10^{108}$. The total number of solutions can be reduced by applying different sets of constraints on the UC problem. The computational complexity in UC problem can be further exacerbated by incorporating uncertainties that represent the realistic operation of power system units such as uncertainty in load variations. In such cases the scenarios for possible solutions might be even more higher ~\cite{Uncertainty} \cite{SUC}. 

Quantum computing is an emerging computing technology that has potential to tackle such complex power system problems which are otherwise intractable for large-scale systems using classical computers. The 2024 Quantum Information Science Applications Roadmap published by the DOE\cite{doeQIS2024} mentions smart allocation of resources in power grids as one of the use cases for quantum computing based optimization methods. 
UC is one of the power system optimization problems that has been researched using multiple quantum computing algorithms~\cite{koretsky2021qaoa_uc},~\cite{Ajagekar_2019},~\cite{Peng_zhang_1},~\cite{Peng_zhang_2}. In~\cite{Ajagekar_2019}, the unit commitment problem is reformulated into a quadratic unconstrained binary optimization (QUBO) and then solved on the D-Wave processor. In ~\cite{UC_D_wave_250}, the UC problem is reformulated as a QUBO problem using D-wave's Binary Quadratic model (BQM) that employs logarithm discretization strategy. A Quantum Approximation Optimization Algorithm (QAOA) is used to solve the unit commitment problem in ~\cite{koretsky2021qaoa_uc}. Quantum Alternate Direction Method of Multipliers (Q-ADMM) methods were utilized in~\cite{Peng_zhang_1}, ~\cite{Peng_zhang_2} to solve the UC problem. The unit commitment problems were also solved using Quantum Surrogate Lagrangian Relaxation (QSLR) method in~\cite{Peng_zhang_3}. In all the aforementioned research works, either the accuracy or scalability or both metrics can be improved so that quantum algorithms can be applied to solve large-scale power system optimization problems such as UC with greater efficiency. 

In this paper, we introduce a Bender's-type heuristic for solving the UC 
problem that has very less mean approximation error rates and to systems with more than 20 generating units for various time-periods. Our heuristic produces an approximate solution to the UC problem in three stages: first it leverages a variational quantum algorithm (VQA) to produce a collection of unit commitment vectors capable of satisfying the power demand with the lowest possible minimum operating costs; then, the heuristic leverages the 
classical sequential least squares programming (SLSQP) routine to find the optimal power level corresponding to a predetermined number of candidate assignment vectors; and finally, it outputs the unit commitment and power level combination achieving the lowest operating cost. The hybrid quantum algorithm is tested for $3$ different use-cases: a) $3$-generator use-case with $4$ time-periods; b) $10$-generator use-case with $24$ time-periods; and c) $26$-generator use-case with $24$ hour time-period. Furthermore, in case of $26$-generator use-case inference on pre-optimized instances along with an end-to-end iterative solution for one of the problem instances on IonQ's Forte quantum computer are demonstrated.  

\section{Unit commitment as a mixed-integer program}
\label{sec:UC}

The UC problem is a mixed-integer program (MIP) that is specified
formally as follows. 

For each generating unit $j = 0, \ldots, N-1$ and each time period $t = 1,
\ldots, T-1$, there is a decision variable $u_j^t \in \{0, 1\}$ that
determines whether unit $j$ is on or off during time period $t$, and there
is a real variable $p_j^t$  such that $p_{j, \min} \leq p_j^t \leq p_{j, \max}$ which determines the power level output by unit $j$ during time period $t$. The
cost of generating power level $p_j^t$ using generator $j$ at time period
$t$ is defined as
\begin{align}
    F_j^t(p) = a_j^t p^2 + b_j^t p + c_j^t.
\end{align}

The objective is to minimize the total cost of generating enough power to
meet a given load $\ell^t$ at each time period. Mathematically, the UC program is written as:

\begin{align}
\begin{aligned}
    \mathrm{minimize}_{u, p} \quad& \sum_{t = 0}^{T-1} \sum_{j=0}^{N-1} u_j^t \, F_j^t(p_j^t), \\
    \text{subject to} \quad& \sum_j p^{t}_j = \ell^t, \quad \text{for} \quad t = 0, \ldots, T-1, \\
    &u_j^t \cdot p_{j, \min} \leq p_j^t \leq u_j^t \cdot p_{j, \max}, \quad \\
    & \quad    \forall \quad  j = 0, \ldots, N-1, \quad t = 0, \ldots, T-1 \\
    & u_j^t \in \{0, 1\}, \quad p_j^t \in \mathbb{R}.
\end{aligned}
\end{align}

Note that this problem is separable, so it is solved one time period at a
time. In other words, for each $t = 1, \ldots, T$, the following optimization problem is solved

\begin{align}
\label{eq:hourly-prob}
\begin{aligned}
    \mathrm{minimize}\quad& \sum_{j=0}^{N-1} u_j^t \, F_j^t(p_j^t), \\
    \text{subject to} \quad& \sum_j p^{t}_j = \ell^t, \\
    &u_j^t \cdot p_{j, \min} \leq p_j^t \leq u_j^t \cdot p_{j, \max}, \\
    & u_j^t \in \{0, 1\}, \quad p_j^t \in \mathbb{R}.
\end{aligned}
\end{align}

In what follows, we refer to the problem defined by Equation~\eqref{eq:hourly-prob} as an \textit{hourly UC problem} and drop the superscript $t$.

\section{Methodology}
\label{sec:methodology}

To tackle the challenge of solving the UC problem, a Bender's-type 
decomposition is introduced making each hourly MIP amenable to a hybrid quantum-classical 
algorithm. 

At a high-level, our decomposition method approximately solves each hourly MIP
in two steps: first it decides which units to turn ``on,'' and then it
computes the optimal power level for each unit it has decided to run. The heuristic first uses a quantum computer to find feasible unit commitment
vectors with low minimum generating costs; it then uses the classical SLSQP
routine to solve a number of residual Quadratic Programs (QPs), corresponding
to fixed unit commitment vectors to find optimal power levels. Finally, our method determines the best unit commitment vector by checking the
overall objective value.

Our solver is based on the following methodology: solving each
hourly MIP is equivalent to solving $2^N$ real residual QPs, one for every possible unit commitment vector $\mathbf{u}$. Formally, the following equality is obtained:
\begin{align}
\label{eq:min-min}
\begin{aligned}
    &\min_{u, p} \bigg\{ \sum_{j=0}^{N-1} u_j \, F_j(p_j) \mid \mathbf{p} \in \Omega(\mathbf{u}) \bigg\} \\
    &=
    \min_w \bigg\{ \min_p \big\{\sum_{j=0}^{N-1} w_j \, F_j(p_j) \mid \mathbf{p} \in \Omega(\mathbf{w})\big\} \mid \mathbf{w} \in \{0, 1\}^N \bigg\}.
\end{aligned}
\end{align}
In the last equation,
$\Omega(\mathbf{u}) = \{\mathbf{p} \mid  \mathbf{u}^T \mathbf{p} = \ell, \; u_j p_{j, \min} \leq p_j \leq u_j p_{j, \max} \}$
denotes the feasible polytope corresponding to the unit commitment vector
$\mathbf{u}$. For simplicity of notation, the solution to
the residual QP corresponding to the unit commitment vector $\mathbf{u}$ is denoted by
\begin{align}
    \mathrm{RQP}(\mathbf{u}) = \min_p \bigg\{ \sum_{j=0}^{N-1} u_j \, F_j(p_j) \mid \mathbf{p} \in \Omega(\mathbf{u}) \bigg\}.
\end{align}
With this notation in hand, Equation~\eqref{eq:min-min} can be written as
\begin{align}
\label{eq:min-rqp}
    \min_{u, p} \bigg\{ \sum_{j=0}^{N-1} &u_j \, F_j(p_j) \mid \mathbf{p} \in \Omega(\mathbf{u}) \bigg\} \\
    &=
    \min_w \{ \mathrm{RQP}({\mathbf{w}}) \mid \mathbf{w} \in \{0, 1\}^N \}. \nonumber
\end{align}

Thus, \textit{in principle}, it is possible to solve each hourly MIP by solving $2^N$ real
residual QPs, one for each unit commitment vector. Naturally, solving an exponential number of residual problems to obtain solutions for each hourly MIP is computationally infeasible. The key challenge, therefore, lies in identifying a ``reasonably small'' subset of residual programs that can be used to produce a sufficiently accurate approximation to the global optimum.

Equation~\eqref{eq:min-rqp} forms the basis of our decomposition method: leverage a quantum computer to ``sieve'' the exponential search space of possible unit commitment vectors, leaving a manageable subset of ``candidate'' vectors that can then be processed on a classical machine using established classical methods. Concretely, if $OPT$ denotes the optimal solution to the hourly UC~\eqref{eq:hourly-prob}, our decomposition method produces the following approximation:
\begin{align}
\begin{aligned}
    OPT &= \min_w \{ \mathrm{RQP}({\mathbf{w}}) \mid \mathbf{w} \in \{0, 1\}^N \} \\
    &\approx \min_w \{ \mathrm{RQP}({\mathbf{w}}) \mid \mathbf{w} \in C^* \},
\end{aligned}
\end{align}
where $C^*$ denotes the set of candidate vectors identified with the help of a quantum processor. For the problem instances considered here, with up to $26$ generating units, $C^*$ satisfies $\mathrm{card}(C^*) \leq 128$.

\subsection{Quantum sieve}

We obtain $C^*$ by approximately solving a Hamiltonian energy minimization problem using a Variational Quantum Algorithm (VQA). 

To begin, we turn to setting up a QUBO, which we will later encode as a Hamiltonian energy minimization problem and then solve approximately using a VQA. The QUBO is constructed using the following observations.

\textbf{Observation 1} The cost function $F_j$ is strictly increasing,
so $F_j(p_{j, \min}) \leq F_j(p_j)$ for every feasible $p_j$, if $u_j = 1$.

This implies that
\begin{align}
    c_{\min}(\mathbf{u}) = \sum_j u_j \, F_j(p_{j, \min})
\end{align}
is a lower bound for the cost of operating the units committed by
$\mathbf{u}$. Thus  $\mathbf{u}$ is called a \textit{solution candidate} if $c_{\min}(\mathbf{u}) \leq OPT$.
Notice that if a vector $\mathbf{u}$ is not a solution candidate,
it cannot solve the hourly MIP, since
\begin{align}
    OPT < c_{\min}(\mathbf{u}) \leq \sum_j u_j F_j(p_j).
\end{align}
Thus it follows that assignment vectors with the lowest
$c_{\min}$ values are the most likely to be solution candidates.

\textbf{Observation 2} Feasible assignment vectors can be easily filtered by checking if $\mathbf{u}^T \mathbf{p}_{\max} \geq \ell$. 

This follows immediately by definition: $\mathbf{u}$ is said to be \textit{feasible} if $\Omega(\mathbf{u}) = \{\mathbf{p} \mid  \mathbf{u}^T \mathbf{p} = \ell, \; u_j p_{j, \min} \leq p_j \leq u_j p_{j, \max} \}$ is nonempty, and the latter is true only if $\mathbf{u}^T \mathbf{p}_{\max} \geq \ell$.

Combined, the two observations imply that feasible assignment vectors with the lowest $c_{\min}$ values are most likely to be solution
candidates. In other words, we seek approximate solutions to the following binary LP:
\begin{align}
\label{eq:lp-qubo}
\begin{aligned}
    \mathrm{minimize}\quad& c_{\min}(\mathbf{u}), \\
    \text{subject to}\quad& \mathbf{u}^T \mathbf{p}_{\max} \geq \ell.
\end{aligned}
\end{align}

In practice, we encode this binary LP as a QUBO with objective function given by a linear combination of $c_{\min}$ and the penalty term
\begin{align}
    P(\mathbf{u}) = \mathrm{erf}\big((\ell - \mathbf{u}^T \mathbf{p}_{\max})^+ \big).
\end{align}
In the last equality, $\mathrm{erf}$ denotes the Gaussian error function and $(\cdot)^+ = \max(\cdot, 0)$, so $(\ell - \mathbf{u}^T \mathbf{p}_{\max})^+$ denotes the positive part of the constraint violation. Concretely, the QUBO objective is given by
\begin{align}\label{eq:hourly-qubo}
    Q(\mathbf{u}) = c_{\min}(\mathbf{u}) + \lambda \, P(\mathbf{u}),
\end{align}
with $\lambda>0$ denoting a tunable hyper-parameter defining the trade-off between minimizing minimal operating costs and satisfying the requested power load.

Whereas standard techniques typically introduce slack variables when converting a problem with inequality constraints into a QUBO, instead we leverage the method introduced in \cite{ionq2025airbus}, which computes penalty terms as a function of constraint violations as it lazily evaluates the map $\mathbf{u} \mapsto Q(\mathbf{u})$. This method obviates the need for ancilla qubits corresponding to the slack variables when solving the QUBO using a VQA.

The QUBO defined by Equation~\eqref{eq:hourly-qubo} is then encoded as a Hamiltonian energy minimization problem by formulating a Hamiltonian $H_Q$ on $N$ qubits such that $H_Q \ket{\mathbf{u}} = Q(\mathbf{u}) \ket{\mathbf{u}}$. A decomposition of $H_Q$ as a linear combination of Pauli operators is not needed; as explained in \cite{ionq2025airbus}, the eigenvalues $Q(\mathbf{u})$ of $H_Q$ are evaluated \textit{lazily} on the classical device as the hybrid quantum-classical ensemble executes the VQA to approximately minimize the energy of $H_Q$.

\begin{figure}[!t]
\centering
\begin{adjustbox}{max width=\columnwidth}
\begin{quantikz}[
  row sep=0.25cm, 
]
\lstick{$\ket{u_1^*}_{\mathrm{relax}}$} & \gate[wires=6, style={minimum width=1.2cm}]{U_E(\boldsymbol{\gamma}^{(1)})}
                   & \gate[wires=6, style={minimum width=0.4cm}]{U_M(\beta_1)}
                   & \gate[wires=6, style={minimum width=1.2cm}]{U_E(\boldsymbol{\gamma}^{(2)})}
                   & \gate[wires=6, style={minimum width=0.4cm}]{U_M(\beta_2)} & \qw \\
\lstick{$\ket{u_2^*}_{\mathrm{relax}}$} & \push{\rule{0pt}{0.9em}} & & & & \qw \\
\lstick{$\ket{u_3^*}_{\mathrm{relax}}$} & \push{\rule{0pt}{0.9em}} & & & & \qw \\
\lstick{$\ket{u_4^*}_{\mathrm{relax}}$} & \push{\rule{0pt}{0.9em}} & & & & \qw \\
\lstick{$\ket{u_5^*}_{\mathrm{relax}}$} & \push{\rule{0pt}{0.9em}} & & & & \qw \\
\lstick{$\ket{u_6^*}_{\mathrm{relax}}$} &                          & & & & \qw \\
\end{quantikz}
\end{adjustbox}
\caption{Illustration of our layered alternating ansatz $\ket{\psi(\boldsymbol{\theta})}$ on $N = 6$ qubits, with $2$ layers, and $\boldsymbol{\theta} = (\boldsymbol{\gamma}^{(1)}, \beta_1, \boldsymbol{\gamma}^{(2)}, \beta_2)$. Here $U_E$ denotes the entangling block and $U_M$ denotes the mixer. The entangling block has the structure of the Butterfly Ansatz, as in \cite{Cherrat2024quantumvision}, constructed with parameteric $ZY$ gates, and the mixer is constructed so its ground state $\ket{\mathbf{u}^*}_{\mathrm{relax}}$ encodes the solution to the semi-definite relaxation of Problem~\eqref{eq:lp-qubo}. We note that each $\boldsymbol{\gamma}^{(\ell)}$ is a parameter vector with $O(\log N)$ parameters, one for each ``layer'' in the Butterfly ansatz.}
\label{fig:ansatz}
\end{figure}
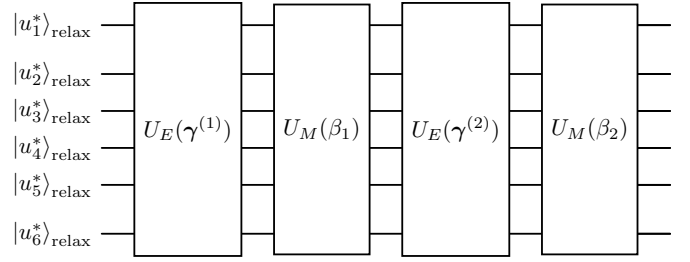

\begin{figure*}[!h]
\centering

\begin{subfigure}[b]{0.48\textwidth}
    \centering
    \includegraphics[width=\textwidth]{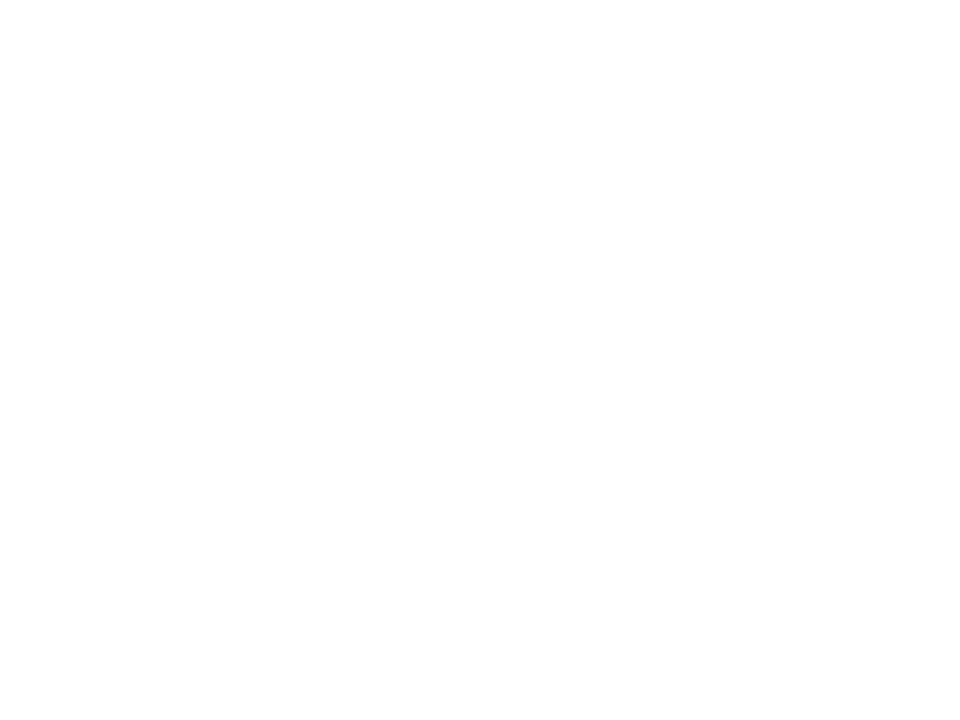}
\end{subfigure}
\hfill
\begin{subfigure}[b]{0.48\textwidth}
    \centering
    \includegraphics[width=\textwidth]{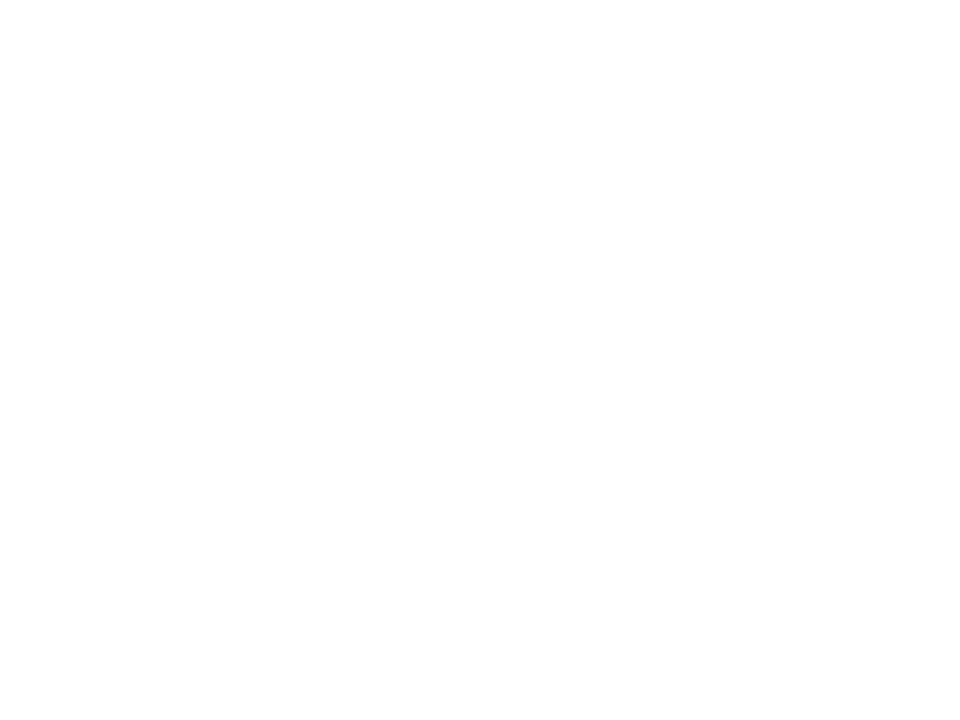}
\end{subfigure}

\caption{(Left panel) Mean approximation error across $24$ time periods in our $10$-unit problem. (Right panel) Mean number of COBYLA iterations required for convergence to a tolerance of $10^{-6}$ across $24$ time periods in our $10$-unit UC instance. In both cases, the results are averaged over $7$ independent trials. Each trial consisted of a noiseless quantum simulation of the VQA described in Section~\ref{sec:methodology}, with $\lambda = 450,000$ and solving at most $128$ residual QPs for each hourly problem. Combined, these figures illustrate the tradeoff between the total cost of the ansatz parameter optimization routine and the achievable approximation error.}
\label{fig:complexity-error-tradeoff}
\end{figure*}

\begin{figure*}[!h]
\centering

\begin{subfigure}[b]{0.48\textwidth}
    \centering
    \includegraphics[width=\textwidth]{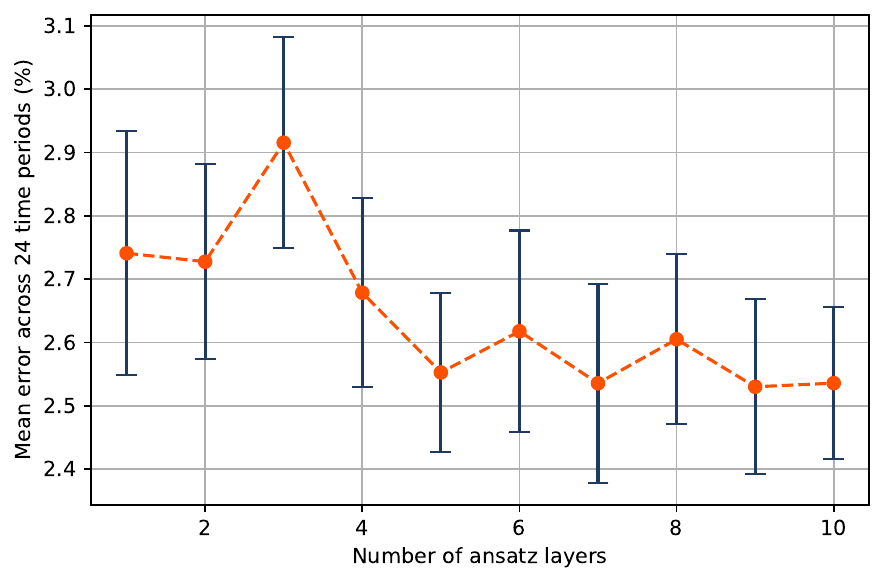}
\end{subfigure}
\hfill
\begin{subfigure}[b]{0.48\textwidth}
    \centering
    \includegraphics[width=\textwidth]{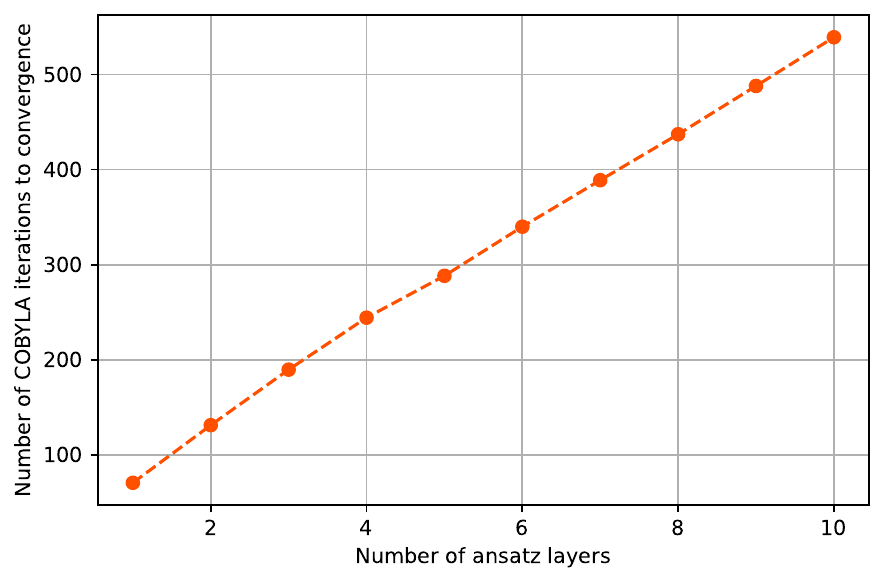}
\end{subfigure}

\caption{(Left panel) Mean approximation error across $24$ time periods in our $26$-unit problem. (Right panel) Mean number of COBYLA iterations required for convergence to a tolerance of $10^{-6}$ across $24$ time periods in our $26$-unit UC instance. In both cases, the results are averaged over $3$ independent trials. Each trial consisted of a noiseless quantum simulation of the VQA described in Section~\ref{sec:methodology}, with $\lambda = 700,000$ and solving at most $128$ residual QPs for each hourly problem. Combined, these figures illustrate the tradeoff between the total cost of the ansatz parameter optimization routine and the achievable approximation error.}
\label{fig:complexity-error-tradeoff-26-units}
\end{figure*}

Concretely, our VQA solves approximately the following non-convex optimization problem:
\begin{align}
    \mathrm{minimize}_{\boldsymbol{\theta}} \bra{\psi(\boldsymbol{\theta})} H_Q \ket{\psi(\boldsymbol{\theta})}
\end{align}
with $\ket{\psi(\boldsymbol{\theta})}$ denoting the parametrized quantum state defined by the variational circuit illustrated in Figure~\ref{fig:ansatz}.
The design for our quantum ansatz is motivated by the following considerations. As the number of parameters increases, the number of iterations required for convergence increases; that is, the ansatz becomes harder to ``train.'' Conversely, as the number of parameters increases, it becomes more likely that there exists a set of parameter values $\boldsymbol{\theta}^*$ such that $\ket{\psi({\boldsymbol{\theta}^*)}}$ is the ground state of $H_Q$; in other words, the ansatz has greater ``expressivity.'' However, as the number of gates in the ansatz increases, the quantum computational complexity increases and computations become more expensive. In addition, as the number of gates increases, the effect of QPU noise becomes more significant. Figures~\ref{fig:complexity-error-tradeoff} and~\ref{fig:complexity-error-tradeoff-26-units} illustrate the trade-off between the number of circuit parameters and the number of iterations required for training under ideal quantum simulation.

The ansatz used here is designed to strike a balance between trainability, expressivity, gate count, and overall depth. We introduce a layered ansatz that alternates between an ``entangling'' block and a ``mixer'' block, much like the Quantum Alternating Operator Ansatz introduced in \cite{qaoa_ansatz}. 

Our entangling block has the structure of the Butterfly Ansatz introduced in \cite{Cherrat2024quantumvision}, with the parametric Reconfigurable Beam Splitter (RBS) gates replaced by parametric $ZY$ gates. The Butterfly Ansatz layout is leveraged for its demonstrated trainability and expressivity at only $O(N \log N)$ gates and $O(\log N)$ depth \cite{Cherrat2024quantumvision}. The $ZY$ gates are used because they can be implemented using a single two-qubit gate on IonQ's processors. It is worth noting that the all-to-all connectivity of qubits in IonQ systems enables such an implementation for qubits with non-adjacent indices.

Our mixer block is defined using the warm-starting procedure introduced in \cite{Tate2023warmstartedqaoa}. In short, this procedure defines a mixer as a separable Hamiltonian whose ground state $\ket{\mathbf{u}^*}_{\mathrm{relax}}$ encodes the solution of the semi-definite relaxation of Problem~\eqref{eq:lp-qubo}. 

Moreover, our ansatz sets the qubits to the initial state $\ket{\mathbf{u}^*}_{\mathrm{relax}}$ using a layer of parametrized single-qubit rotations, as described in \cite{Tate2023warmstartedqaoa}.

Upon convergence of the VQA, the optimized quantum state is sampled to obtain a collection of feasible assignment vectors with low minimum operating costs--these have the highest chance of being solution candidates. The samples are ranked according to their $c_{\min}$ value and we keep only a fixed number of samples, which is chosen ahead of time. The remaining subset is returned as $C^*$.

\subsection{Classical refinement}

Following the sampling of the optimized quantum circuit, the residual QP corresponding to each feasible unit commitment vector in $C^*$ is solved to obtain a collection of $\mathrm{RQP}(\mathbf{u})$ values. In this study we solve the residual problems using the SLSQP routine; however, other classical methods can be leveraged. To conclude, the best assignment is reported as
\begin{align}
    \mathrm{argmin}_{\mathbf{u}} \{\mathrm{RQP}(\mathbf{u}) \mid \mathbf{u} \in C^*\}.
\end{align}

\section{Results}
\label{sec:results}

In this section we report the performance of our algorithm on three UC instances with varying numbers of generators and time periods. In particular, we report quantum simulation results on a $3$-generator instance with $4$ time periods, a $10$-generator instance with $24$ time periods, and a $26$-generator instance with $24$ time periods in Section~\ref{sec:simulator}, and we report approximation results for selected time periods of the $10$- and $26$-unit instances produced using IonQ Forte in Section~\ref{sec: quantum Computer}. Further results on the 3-generator use-case can be found in the Appendix. The data for the generator units of the UC instances considered here is reported in Tables~\ref{tab:3-units-data},~\ref{tab:10-units-data}, and~\ref{tab:26-units-data}. These data are taken from~\cite{Wood},~\cite{ehg_uc_10_units}, and~\cite{nff_uc_26_units}, respectively.

\begin{table}[hbt]
\centering
\caption{Generator specs for our $3$-unit grid UC instance. The data are taken from Ref.~\cite{Wood}.}
\label{tab:3-units-data}
\begin{tabular}{lrrrrr}
\toprule
 & $p_{j, \min}$ & $p_{j, \max}$ & $c_j$ & $b_j$ & $a_j$ \\
Unit & (MW) & (MW) &  &  &  \\
\midrule
0 & 100 & 600 & 500 & 10 & 0.002 \\
1 & 100 & 400 & 300 & 8 & 0.0025 \\
2 & 50 & 200 & 100 & 6 & 0.005 \\
\bottomrule
\end{tabular}
\end{table}

\begin{table}[hbt]
\centering
\caption{Generator specs for our $10$-unit grid UC instance. The data are taken from Ref.~\citen{ehg_uc_10_units}.}
\label{tab:10-units-data}
\begin{tabular}{lrrrrr}
\toprule
 & $p_{j, \min}$ & $p_{j, \max}$ & $c_j$ & $b_j$ & $a_j$ \\
Unit & (MW) & (MW) &  &  &  \\
\midrule
0 & 150 & 455 & 1000 & 16.19 & 0.00048 \\
1 & 150 & 455 & 970 & 17.26 & 0.00031 \\
2 & 20 & 130 & 700 & 16.6 & 0.00200 \\
3 & 20 & 130 & 680 & 16.5 & 0.00211 \\
4 & 25 & 162 & 450 & 19.7 & 0.00398 \\
5 & 20 & 80 & 370 & 22.26 & 0.00712 \\
6 & 25 & 85 & 480 & 27.74 & 0.00079 \\
7 & 10 & 55 & 660 & 25.92 & 0.00413 \\
8 & 10 & 55 & 665 & 27.27 & 0.00222 \\
9 & 10 & 55 & 670 & 27.79 & 0.00173 \\
\bottomrule
\end{tabular}
\end{table}

\begin{table}[hbt]
\centering
\caption{Generator specs for our $26$-unit grid UC instance. The data are taken from Ref.~\citen{nff_uc_26_units}.}
\label{tab:26-units-data}
\begin{tabular}{lrrrrr}
\toprule
 & $p_{j, \min}$ & $p_{j, \max}$ & $c_j$ & $b_j$ & $a_j$ \\
Unit & (MW)  & (MW) &  &  &  \\
\midrule
0 & 2.4 & 12 & 24.3891 & 25.55 & 0.02533 \\
1 & 2.4 & 12 & 24.411 & 25.68 & 0.02649 \\
2 & 2.4 & 12 & 24.6382 & 25.8 & 0.02801 \\
3 & 2.4 & 12 & 24.7605 & 25.93 & 0.02842 \\
4 & 2.4 & 12 & 24.8882 & 26.06 & 0.02855 \\
5 & 4 & 20 & 117.755 & 37.55 & 0.01199 \\
6 & 4 & 20 & 118.108 & 37.66 & 0.01261 \\
7 & 4 & 20 & 118.458 & 37.78 & 0.01359 \\
8 & 4 & 20 & 118.821 & 37.89 & 0.01433 \\
9 & 15.2 & 76 & 81.1364 & 13.33 & 0.00876 \\
10 & 15.2 & 76 & 81.298 & 13.36 & 0.00895 \\
11 & 15.2 & 76 & 81.4641 & 13.38 & 0.0091 \\
12 & 15.2 & 76 & 81.6259 & 13.41 & 0.00932 \\
13 & 25 & 100 & 217.895 & 18 & 0.00623 \\
14 & 25 & 100 & 218.335 & 18.1 & 0.00612 \\
15 & 25 & 100 & 218.775 & 18.2 & 0.00598 \\
16 & 54.25 & 155 & 142.735 & 10.69 & 0.00463 \\
17 & 54.25 & 155 & 143.029 & 10.72 & 0.00473 \\
18 & 54.25 & 155 & 143.318 & 10.74 & 0.00481 \\
19 & 54.25 & 155 & 143.597 & 10.76 & 0.00487 \\
20 & 68.95 & 197 & 259.131 & 23 & 0.00259 \\
21 & 68.95 & 197 & 259.649 & 23.1 & 0.0026 \\
22 & 68.95 & 197 & 260.176 & 23.2 & 0.00263 \\
23 & 140 & 350 & 177.058 & 10.86 & 0.00153 \\
24 & 100 & 400 & 310.002 & 7.49 & 0.00194 \\
25 & 100 & 400 & 311.91 & 7.5 & 0.00195 \\
\bottomrule
\end{tabular}
\end{table}

\subsection{Quantum simulations}
\label{sec:simulator}

We conducted various runs of the VQA explained in Section~\ref{sec:methodology} on a quantum simulator. In particular, we leveraged Google's \texttt{qsim} state vector simulator in conjunction with Scipy's implementation of the Constrained Optimization by Linear Approximation (COBYLA) algorithm \cite{powell1994cobyla} to tune our variational quantum ansatz. Over the course of the parameter optimization routine, we used $512$ shots to sample the parametrized quantum state at each COBYLA iteration. We used Scipy's default stopping criterion to halt the optimization, and in each case we initialized circuit parameters as $\boldsymbol{\theta} = 0$.

Then we sampled the optimized quantum state $\ket{\psi(\theta^*)}$ resulting from the VQA using $5,000$ shots and solved at most $128$ residual QPs corresponding to feasible assignment vectors sampled from the optimized quantum state. For this step, we choose the sampled feasible assignment vectors with the lowest minimum operating cost, i.e., those belonging to $C^*$.

Our main results are summarized in Tables~\ref{tab:3-units-apx},~\ref{tab:sim-10-units}, and~\ref{tab:sim-26-units}. Our method (trivially) produces the optimal solution to the $3$-unit problem, reported in Table~\ref{tab:3-units-apx}, in every case, so we focus on the $10$- and $26$-unit problems.

\begin{figure*}[hbt]
\centering

\begin{subfigure}[b]{0.48\textwidth}
    \centering
    \includegraphics[width=\textwidth]{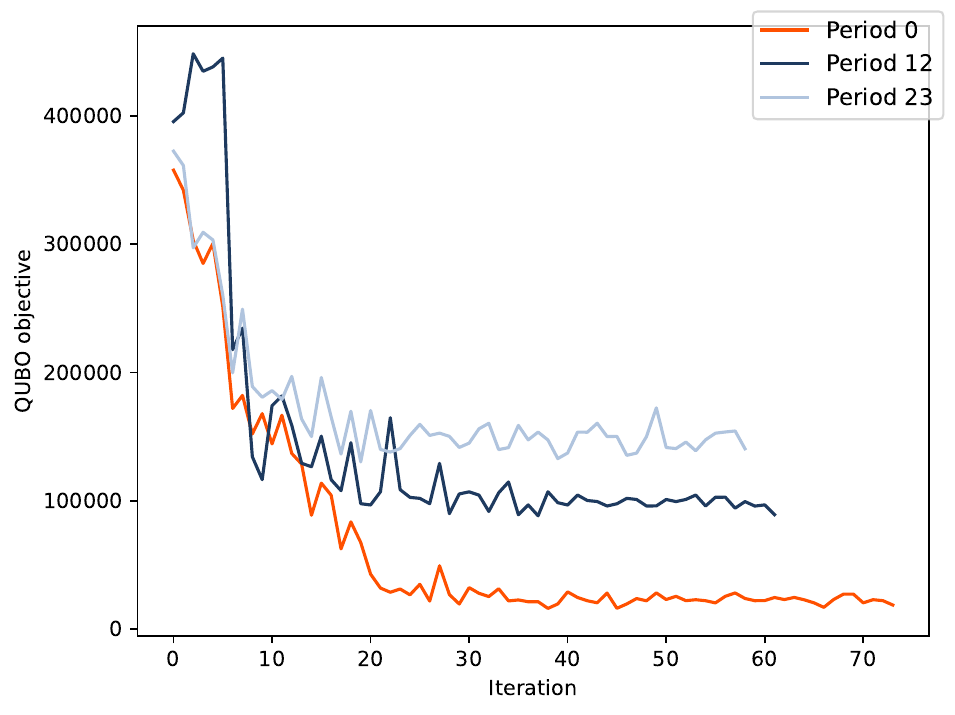}
\end{subfigure}
\hfill
\begin{subfigure}[b]{0.48\textwidth}
    \centering
    \includegraphics[width=\textwidth]{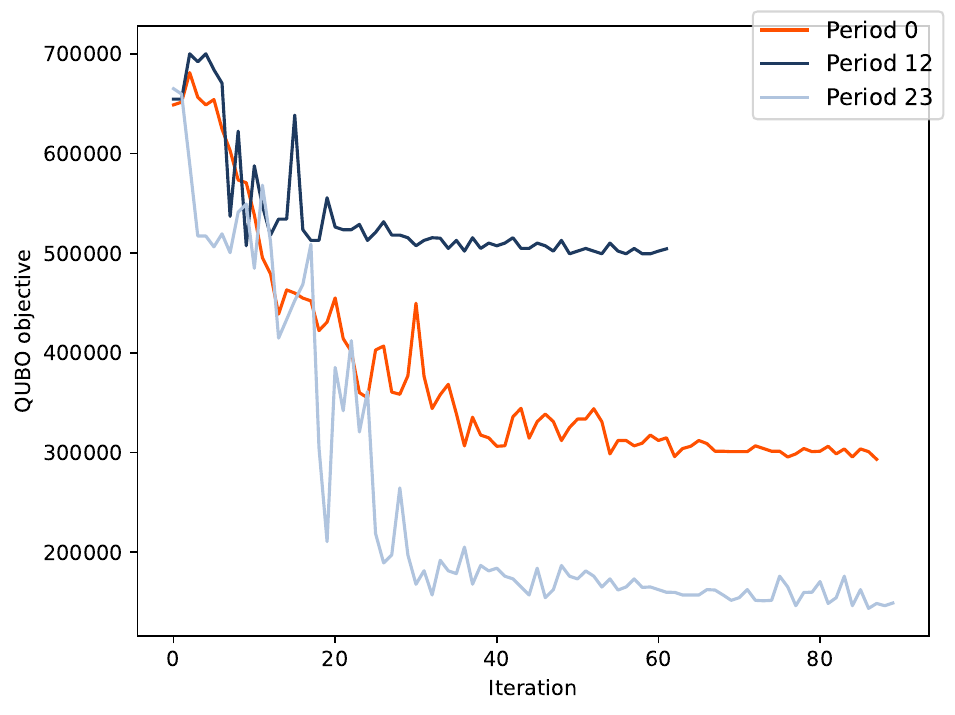}
\end{subfigure}
\caption{Simulated convergence of our VQA when solving our for selected power loads. We used our layered ansatz with a single layer and $512$ shots per iteration. All parameters were initialized to zero. (Left panel) Hourly problems with $10$ generating units. (Right panel) Hourly problems with $26$ generating units.}
\label{fig:vqa-sims}
\end{figure*}

To approximately solve these UC instances, we executed our VQA with varying numbers of layers in our quantum ansatz. For each of the $24$ time periods in the $10$-unit problem, we ran $7$ independent trials to obtain robust results; similarly, we ran $3$ independent trials for each period in the $26$-unit problem. Tables~\ref{tab:sim-10-units} and~\ref{tab:sim-26-units} summarize the results of these experiments, and Figures~\ref{fig:complexity-error-tradeoff} and~\ref{fig:complexity-error-tradeoff-26-units} illustrate them. In addition, Tables~\ref{tab:10-units-apx} and~\ref{tab:26-units-apx} report sample solutions obtained by our hybrid algorithm, and Figure~\ref{fig:vqa-sims} supplements our results summary by illustrating the simulated convergence of our VQA for selected power loads.  

Our results suggest there is a linear relationship between the number of iterations required for ``training'' and the number of parameters in the ansatz; in particular, a least-squares fit suggests the number of iterations required for convergence on average increases by $8.63$ for each additional parameter. As expected, the approximation error decreases as a function of the number of ansatz layers, but the trends in Figures~\ref{fig:complexity-error-tradeoff} and~\ref{fig:complexity-error-tradeoff-26-units} suggest the marginal utility is decreasing.

Given a linear scaling in the number of iterations required for convergence of our VQA, it follows that the quantum computational complexity our method is $O(L^2 \log N)$ Circuit Layer Operations (CLOPs), with $L$ denoting the number of ansatz layers. This complexity result follows because our ansatz has depth $O(L \log N)$, there are $O(L \log N)$ parameters in our ansatz, and $(\log N)^2$ is $O(\log N)$. In terms of two-qubit gates (TQG), the complexity of our VQA is $O(L^2 N \log N)$. Given the decreasing marginal returns on the ansatz layers, we do not expect $L$ scale indefinitely. A deeper understanding of the relationship between $L$ and the approximation error is needed, especially as $N$ grows to industrially relevant scales.

\begin{table}[!h]
\centering
\caption{Approximate solution for 3-unit grid problem at varying loads, computed using our novel hybrid quantum-classical heuristic. This solution achieves the global optimum; that is, our approximation error is exactly $0\%$.}
\label{tab:3-units-apx}
\begin{tabular}{lllllrrrr}
\toprule
Period & Load & \multicolumn{3}{c}{Assignment} & \multicolumn{3}{c}{Power level} & Total cost \\
$t$ & $\ell$ & $u_0$ & $u_1$ & $u_2$ & $p_0$ & $p_1$ & $p_2$ &  \\
\midrule
0 & 170 & 0 & 0 & 1 & 0 & 0 & 170 & 1264.5 \\
1 & 520 & 0 & 1 & 1 & 0 & 320 & 200 & 4616 \\
2 & 1100 & 1 & 1 & 1 & 500 & 400 & 200 & 11400 \\
3 & 330 & 0 & 1 & 1 & 0 & 130 & 200 & 2882.25 \\
\bottomrule
\end{tabular}
\end{table}

\begin{table}[hbt]
    \caption{Simulated results for solving our $10$-unit, $24$-hour, UC problem. The approximation error in the rightmost column is computed by comparing the approximation produced by our quantum-classical heuristic against the global optimum computed by CPLEX~\cite{cplex2009v12}. The number of iterations and the approximation error are averaged across the $24$ time periods with power load profile described in Table~\ref{tab:10-units-apx}, and across $7$ independent trials for each period. The results summarized in this table are illustrated in Figure~\ref{fig:complexity-error-tradeoff}.}
    \centering
\begin{tabular}{cccccc}
\toprule
Ansatz layers & TQG & Params & Depth & Iterations & Error (\%) \\
\midrule
1 & 15 & 5 & 16 & 59.14 & 1.78 \\
2 & 30 & 10 & 31 & 103.63 & 1.18 \\
3 & 45 & 15 & 46 & 148.23 & 0.95 \\
4 & 60 & 20 & 61 & 194.07 & 1.03 \\
5 & 75 & 25 & 76 & 238.61 & 0.75 \\
6 & 90 & 30 & 91 & 277.00 & 0.74 \\
7 & 105 & 35 & 106 & 324.45 & 0.73 \\
8 & 120 & 40 & 121 & 366.80 & 0.72 \\
9 & 135 & 45 & 136 & 407.51 & 0.55 \\
10 & 150 & 50 & 151 & 449.38 & 0.72 \\
\bottomrule
\end{tabular}
    \label{tab:sim-10-units}
\end{table}

\begin{table}[hbt]
    \centering
        \caption{Simulated results for solving our $26$-unit, $24$-hour, UC problem. The approximation error in the rightmost column is computed by comparing the approximation produced by our quantum-classical heuristic against the global optimum computed by CPLEX~\cite{cplex2009v12}. The number of iterations and the approximation error are averaged across the $24$ time periods power load profile described in Table~\ref{tab:26-units-apx}, and $3$ independent trials for each period. The results summarized in this table are illustrated in Figure~\ref{fig:complexity-error-tradeoff-26-units}.}
\begin{tabular}{cccccc}
\toprule
Ansatz layers & TQG & Params & Depth & Iterations & Error (\%) \\
\midrule
1 & 57 & 6 & 19 & 70.67 & 2.74 \\
2 & 114 & 12 & 37 & 131.25 & 2.73 \\
3 & 171 & 18 & 55 & 189.65 & 2.92 \\
4 & 228 & 24 & 73 & 244.29 & 2.68 \\
5 & 285 & 30 & 91 & 288.31 & 2.55 \\
6 & 342 & 36 & 109 & 339.93 & 2.62 \\
7 & 399 & 42 & 127 & 388.82 & 2.54 \\
8 & 456 & 48 & 145 & 437.11 & 2.60 \\
9 & 513 & 54 & 163 & 487.89 & 2.53 \\
10 & 570 & 60 & 181 & 539.08 & 2.54 \\
\bottomrule
\end{tabular}
    \label{tab:sim-26-units}
\end{table}

\subsection{Execution on IonQ QPUs}
\label{sec: quantum Computer}

In this section we report the performance of our hybrid algorithm on the IonQ Forte QPU, which has demonstrated performance at the level of 36 algorithmic qubits \cite{IonQ_blog, IonQ_Aria, IonQ_Forte, lubinski2023application}. Additional figures illustrating the convergence of our VQA when solving the $3$-unit problem are provided in the Appendix. We use these experiments mainly to validate the our simulated results and to study the effect of QPU noise on the quality of our approximations. Overall, we observe qualitative agreement between our VQA simulations and our hardware executions.

\begin{figure}
    \centering
    \includegraphics[width=0.95\linewidth]{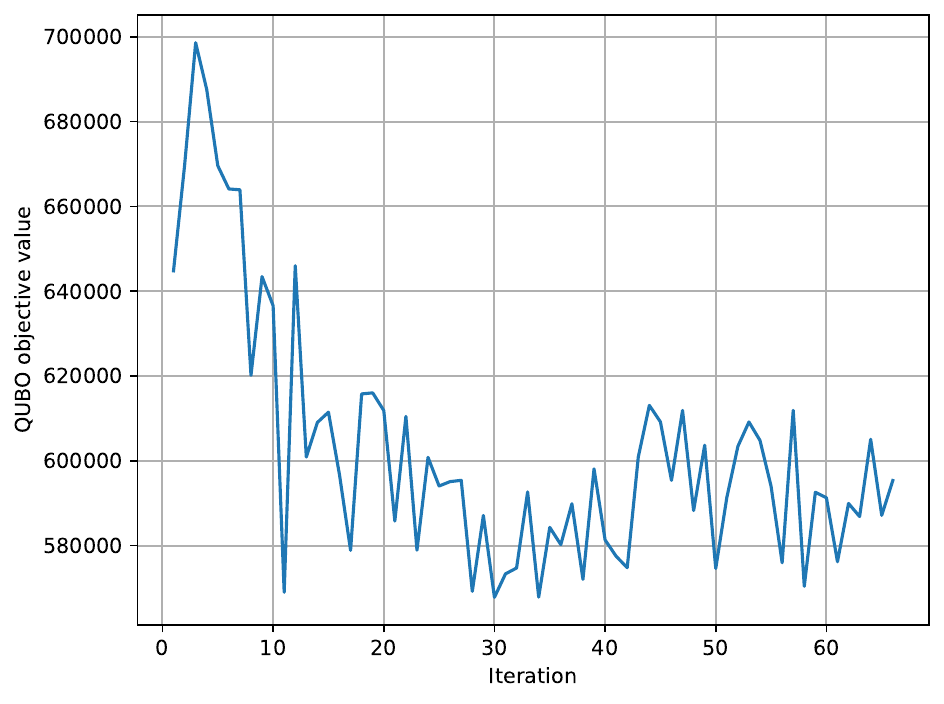}
    \caption{Convergence of our VQA running on IonQ Forte as it approximately solves an hourly problem with $26$ generating units. For this experiment we used our layered ansatz with a single layer and $512$ shots per iteration. All parameters were initialized to zero.}
    \label{fig:26-generator-period-23}
\end{figure}

\begin{figure*}[hbt]
    \centering
    \includegraphics[width=0.95\linewidth]{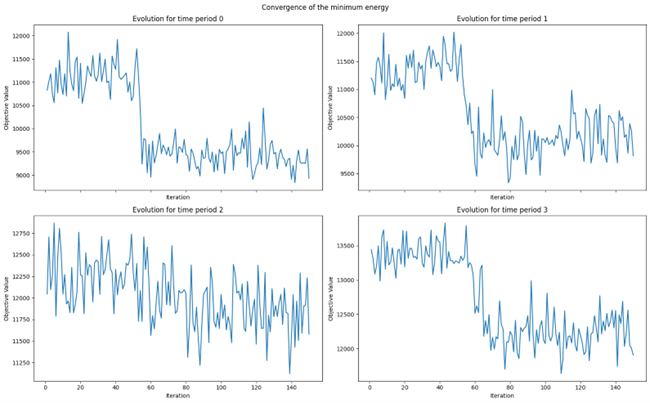}
    \caption{Convergence of our VQA running on IonQ Forte as it approximately solves an hourly problem with $10$ generating units. For this experiment we used our layered ansatz with a single layer and $1,000$ shots per iteration. For this experiment we used a brickwork layout ansatz with $4$ layers of parameterized RY gates and we initialized the parameters by drawing uniformly at random in the interval $[-2\pi, 2\pi]$.}
    \label{fig:10-generator 4-hour Forte}
\end{figure*}

Using IonQ Forte, we executed our VQA to approximately solve a selection of time periods of our $10$- and $26$-unit UC instances. As in our quantum simulations, we used Scipy's COBYLA implementation with default convergence settings to optimize our ansatz parameters. The results for our $10$ unit instance were obtained using an earlier ansatz design, with parametrized $ZY$ gates placed in a brickwork fashion, and we used $1,000$ shots per iteration. For the $26$-unit instance, we used the layered ansatz introduced in Section~\ref{sec:methodology} with a single layer, and we used $512$ shots per COBYLA iteration, as in the simulation experiments. Figure~\ref{fig:10-generator 4-hour Forte} illustrates the convergence of our VQA for select periods of our $10$-unit instance, and Figure~\ref{fig:26-generator-period-23} illustrates convergence of our VQA as it minimizes a $26$-unit hourly problem. These hourly problem instances are defined by the power loads reported in Tables~\ref{tab:10-units-apx} and~\ref{tab:26-units-apx}.

\begin{figure*}[!h]
    \centering
    \includegraphics[width=0.95\linewidth]{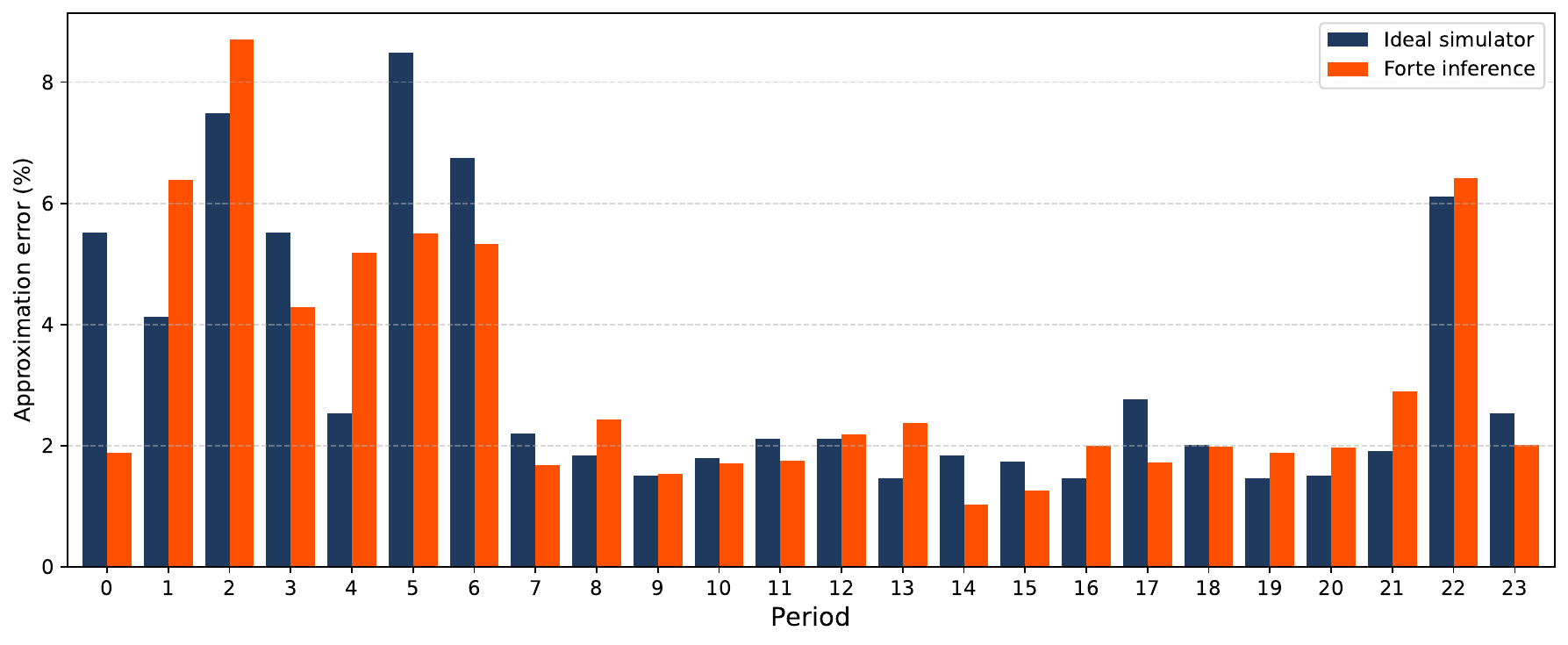}
    \caption{Approximation error obtained when using the quantum simulator and IonQ Forte to sample the optimized quantum state $\ket{\psi(\boldsymbol{\theta^*})}$, described in Section~\ref{sec:methodology}, using optimized ansatz parameters computed by simulating our VQA. For each time period in our $26$-unit problem, we measure $2,000$ samples of the optimized state on each backend and then we solve at most $128$ residual QPs corresponding to the sampled feasible assignment vectors with the lowest minimum operating cost. Here $\ket{\psi}$ denotes our layered ansatz with $1$ gate layer. The average approximation error across time periods when using simulated samples is $3.201\%$, and it is $3.088\%$ when using samples measured by IonQ Forte.}
    \label{fig:forte-inference}
\end{figure*}

In addition, for each of the $24$ periods of our $26$-unit instance, we used IonQ Forte to run ``inference'' jobs: we sampled the optimized quantum ansatze $\ket{\psi(\boldsymbol{\theta^*})}$, using optimized parameters $\boldsymbol{\theta^*}$ obtained on a classical device by simulating our VQA. For this experiment, we measured $2,000$ samples of each optimized state and we used them to solve $128$ residual quadratic programs to produce an approximate solution for each time period. Figure~\ref{fig:forte-inference} illustrates the results: for most time periods, we see good agreement between the values predicted by simulation and those obtained using samples measured on IonQ Forte.

\section{Conclusions and Outlook}
\label{sec:next-steps}

To conclude, a vision for future developments is outlined to improve the performance and scaling of the algorithm presented here. While we have demonstrated an improvement on the current state of the art in quantum-powered solutions to the unit commitment problem, which was achieved by Koretsky \emph{et al.}\cite{koretsky2021qaoa_uc} when the authors solved a number of problem instances using the 10-unit grid described in Section \ref{sec:simulator} to within $8\%$ of the optimal solution, there is still much that can be done to enhance our methodology. 

A promising direction for future research involves replacing our VQE engine with the recent variational quantum imaginary time evolution (varQITE) introduced in ~\cite{morris2024qite}. A successful implementation of a varQITE-based UC solver requires an adaptation of varQITE so it can handle constrained optimization problems. There are at least three research avenues here:
\begin{enumerate}
    \item Experiment with standard translations that convert constrained quadratic programs into QUBOs, and investigate recent advances in efficient constraint encoding to reduce the number of required ancillary qubits.
    \item Develop a decomposition of the quantum cost function that is amenable to the varQITE formulation.
    \item Consider whether other QITE implementations can be more easily adapted to solve constrained problems.
\end{enumerate}

Aside from this, a different avenue is ripe for exploration: application-specific ansatz design. Given the success of this line of work in other applications, it would be advisable to develop a bespoke ansatze that can more easily model the ground state of the Hamiltonian arising in the UC problem. The successful development of such an ansatz would significantly reduce the number of optimizer iterations required to sample high-quality unit commitment allocation vectors.

\section*{Acknowledgment}
Authors would like to thank Rima Oueid and Alireza Ghassemian for overseeing the project developments and providing guidance.

\section*{Appendix}

In the following pages we provide additional tables and figures to accompany the results presented in Section~\ref{sec:results}.

\begin{table*}[hbt]
\centering
\caption{Approximate solution for 10-unit grid problem at varying loads, computed using our novel hybrid quantum-classical heuristic. The ansatz used to obtain this solution had a single layer of gates.}
\label{tab:10-units-apx}
\scalebox{1.0}{\begin{tabular}{lrlrrrrrrrrrrr}
\toprule
 & Load & Assignment & \multicolumn{10}{c}{Power Level} & Total Cost \\
 &  & $\mathbf{u}$ & $p_0$ & $p_1$ & $p_2$ & $p_3$ & $p_4$ & $p_5$ & $p_6$ & $p_7$ & $p_8$ & $p_9$ &  \\
$t$ &  &  &  &  &  &  &  &  &  &  &  &  &  \\
\midrule
0 & 700 & 1001100000 & 455 & 0 & 0 & 130 & 115 & 0 & 0 & 0 & 0 & 0 & 14094.6 \\
1 & 750 & 1001100100 & 455 & 0 & 0 & 130 & 155 & 0 & 0 & 10 & 0 & 0 & 15845.2 \\
2 & 850 & 1100001000 & 455 & 370 & 0 & 0 & 0 & 0 & 25 & 0 & 0 & 0 & 17038.5 \\
3 & 950 & 1100010000 & 455 & 455 & 0 & 0 & 0 & 40 & 0 & 0 & 0 & 0 & 18625.1 \\
4 & 1000 & 1101000000 & 455 & 415 & 0 & 130 & 0 & 0 & 0 & 0 & 0 & 0 & 19512.8 \\
5 & 1100 & 1111000000 & 455 & 385 & 130 & 130 & 0 & 0 & 0 & 0 & 0 & 0 & 21879.3 \\
6 & 1150 & 1111110000 & 455 & 390 & 130 & 130 & 25 & 20 & 0 & 0 & 0 & 0 & 23729.9 \\
7 & 1200 & 1101100000 & 455 & 455 & 0 & 130 & 160 & 0 & 0 & 0 & 0 & 0 & 23917.8 \\
8 & 1300 & 1111100000 & 455 & 455 & 130 & 130 & 130 & 0 & 0 & 0 & 0 & 0 & 26184 \\
9 & 1400 & 1111110000 & 455 & 455 & 130 & 130 & 162 & 68 & 0 & 0 & 0 & 0 & 28768.2 \\
10 & 1450 & 1111111000 & 455 & 455 & 130 & 130 & 162 & 80 & 38 & 0 & 0 & 0 & 30583.2 \\
11 & 1500 & 1111111010 & 455 & 455 & 130 & 130 & 162 & 80 & 33 & 0 & 55 & 0 & 32615.8 \\
12 & 1400 & 1111110000 & 455 & 455 & 130 & 130 & 162 & 68 & 0 & 0 & 0 & 0 & 28768.2 \\
13 & 1300 & 1111110000 & 455 & 455 & 130 & 130 & 110 & 20 & 0 & 0 & 0 & 0 & 26589 \\
14 & 1200 & 1101100000 & 455 & 455 & 0 & 130 & 160 & 0 & 0 & 0 & 0 & 0 & 23917.8 \\
15 & 1050 & 1100100000 & 455 & 455 & 0 & 0 & 140 & 0 & 0 & 0 & 0 & 0 & 20639.3 \\
16 & 1000 & 1101000010 & 455 & 405 & 0 & 130 & 0 & 0 & 0 & 0 & 10 & 0 & 20275.6 \\
17 & 1100 & 1110010000 & 455 & 455 & 130 & 0 & 0 & 60 & 0 & 0 & 0 & 0 & 21976.3 \\
18 & 1200 & 1111010000 & 455 & 455 & 130 & 130 & 0 & 30 & 0 & 0 & 0 & 0 & 24150 \\
19 & 1400 & 1111110000 & 455 & 455 & 130 & 130 & 162 & 68 & 0 & 0 & 0 & 0 & 28768.2 \\
20 & 1300 & 1111100000 & 455 & 455 & 130 & 130 & 130 & 0 & 0 & 0 & 0 & 0 & 26184 \\
21 & 1100 & 1110100000 & 455 & 455 & 130 & 0 & 60 & 0 & 0 & 0 & 0 & 0 & 21891.4 \\
22 & 900 & 1101100000 & 455 & 290 & 0 & 130 & 25 & 0 & 0 & 0 & 0 & 0 & 18272.9 \\
23 & 800 & 1100100000 & 455 & 320 & 0 & 0 & 25 & 0 & 0 & 0 & 0 & 0 & 15935.8 \\
\bottomrule
\end{tabular}}
\end{table*}

\begin{table*}[hbt]
\centering
\caption{Sample approximate solution for $26$-unit grid problem at varying loads, computed using our hybrid quantum-classical heuristic. The ansatz used to obtain this solution had a single layer of gates.}
\label{tab:26-units-apx}
\scalebox{0.57}{
\begin{tabular}{lccrrrrrrrrrrrrrrrrrrrrrrrrrrr}
\toprule
 & Load & Assignment & \multicolumn{26}{c}{Power Level} & Total Cost \\
 & $\ell$ & $\mathbf{u}$ & $p_{0}$ & $p_{1}$ & $p_{2}$ & $p_{3}$ & $p_{4}$ & $p_{5}$ & $p_{6}$ & $p_{7}$ & $p_{8}$ & $p_{9}$ & $p_{10}$ & $p_{11}$ & $p_{12}$ & $p_{13}$ & $p_{14}$ & $p_{15}$ & $p_{16}$ & $p_{17}$ & $p_{18}$ & $p_{19}$ & $p_{20}$ & $p_{21}$ & $p_{22}$ & $p_{23}$ & $p_{24}$ & $p_{25}$ &  \\
$t$ &  &  &  &  &  &  &  &  &  &  &  &  &  &  &  &  &  &  &  &  &  &  &  &  &  &  &  &  &  \\
\midrule
0 & 1700 & 01000000001010101011000111 & 0 & 2.4 & 0 & 0 & 0 & 0 & 0 & 0 & 0 & 0 & 30.71 & 0 & 26.89 & 0 & 25 & 0 & 155 & 0 & 155 & 155 & 0 & 0 & 0 & 350 & 400 & 400 & 18880.9 \\
1 & 1730 & 01000001111000001111000111 & 0 & 2.4 & 0 & 0 & 0 & 0 & 0 & 4 & 4 & 15.2 & 15.2 & 0 & 0 & 0 & 0 & 0 & 141.94 & 136.7 & 132.21 & 128.35 & 0 & 0 & 0 & 350 & 400 & 400 & 19308.8 \\
2 & 1690 & 01110101000100100111000111 & 0 & 2.4 & 2.4 & 2.4 & 0 & 4 & 0 & 4 & 0 & 0 & 0 & 34.8 & 0 & 0 & 25 & 0 & 0 & 155 & 155 & 155 & 0 & 0 & 0 & 350 & 400 & 400 & 19208 \\
3 & 1700 & 01000000001010101011000111 & 0 & 2.4 & 0 & 0 & 0 & 0 & 0 & 0 & 0 & 0 & 30.71 & 0 & 26.89 & 0 & 25 & 0 & 155 & 0 & 155 & 155 & 0 & 0 & 0 & 350 & 400 & 400 & 18880.9 \\
4 & 1750 & 10100101010010001011000111 & 2.4 & 0 & 2.4 & 0 & 0 & 4 & 0 & 4 & 0 & 65.21 & 0 & 0 & 56.99 & 0 & 0 & 0 & 155 & 0 & 155 & 155 & 0 & 0 & 0 & 350 & 400 & 400 & 19745.1 \\
5 & 1850 & 01010000011100101110000111 & 0 & 2.4 & 0 & 2.4 & 0 & 0 & 0 & 0 & 0 & 71.35 & 68.11 & 65.74 & 0 & 0 & 25 & 0 & 155 & 155 & 155 & 0 & 0 & 0 & 0 & 350 & 400 & 400 & 21118.4 \\
6 & 2000 & 11011000101110110111000111 & 2.4 & 2.4 & 0 & 2.4 & 2.4 & 0 & 0 & 0 & 4 & 0 & 76 & 76 & 76 & 0 & 75 & 68.4 & 0 & 155 & 155 & 155 & 0 & 0 & 0 & 350 & 400 & 400 & 24350.1 \\
7 & 2430 & 01111001011111011111001111 & 0 & 2.4 & 2.4 & 2.4 & 2.4 & 0 & 0 & 4 & 0 & 76 & 76 & 76 & 76 & 100 & 0 & 100 & 155 & 155 & 155 & 155 & 0 & 0 & 142.4 & 350 & 400 & 400 & 32100.4 \\
8 & 2540 & 11100111111111111111001111 & 2.4 & 2.4 & 2.4 & 0 & 0 & 4 & 4 & 4 & 4 & 76 & 76 & 76 & 76 & 100 & 100 & 100 & 155 & 155 & 155 & 155 & 0 & 0 & 142.8 & 350 & 400 & 400 & 34918.3 \\
9 & 2600 & 11001011111111111111001111 & 8.7 & 5.9 & 0 & 0 & 2.4 & 0 & 4 & 4 & 4 & 76 & 76 & 76 & 76 & 100 & 100 & 100 & 155 & 155 & 155 & 155 & 0 & 0 & 197 & 350 & 400 & 400 & 36210.3 \\
10 & 2670 & 11100011111111111111101111 & 2.4 & 2.4 & 2.4 & 0 & 0 & 0 & 4 & 4 & 4 & 76 & 76 & 76 & 76 & 100 & 100 & 100 & 155 & 155 & 155 & 155 & 158.62 & 0 & 118.18 & 350 & 400 & 400 & 38034.6 \\
11 & 2590 & 11110010111111111111010111 & 3.8 & 2.4 & 2.4 & 2.4 & 0 & 0 & 4 & 0 & 4 & 76 & 76 & 76 & 76 & 100 & 100 & 100 & 155 & 155 & 155 & 155 & 0 & 197 & 0 & 350 & 400 & 400 & 35788 \\
12 & 2590 & 11110010111111111111010111 & 3.8 & 2.4 & 2.4 & 2.4 & 0 & 0 & 4 & 0 & 4 & 76 & 76 & 76 & 76 & 100 & 100 & 100 & 155 & 155 & 155 & 155 & 0 & 197 & 0 & 350 & 400 & 400 & 35788 \\
13 & 2550 & 01011111111111111111010111 & 0 & 2.4 & 0 & 2.4 & 2.4 & 4 & 4 & 4 & 4 & 76 & 76 & 76 & 76 & 100 & 100 & 100 & 155 & 155 & 155 & 155 & 0 & 152.8 & 0 & 350 & 400 & 400 & 35143.8 \\
14 & 2620 & 11111001011111101111011111 & 2.4 & 2.4 & 2.4 & 2.4 & 2.4 & 0 & 0 & 4 & 0 & 76 & 76 & 76 & 76 & 100 & 100 & 0 & 155 & 155 & 155 & 155 & 0 & 175.51 & 154.49 & 350 & 400 & 400 & 36861.1 \\
15 & 2650 & 11111110111111111111001111 & 12 & 12 & 12 & 12 & 12 & 9.98 & 5.02 & 0 & 4 & 76 & 76 & 76 & 76 & 100 & 100 & 100 & 155 & 155 & 155 & 155 & 0 & 0 & 197 & 350 & 400 & 400 & 37650.7 \\
16 & 2550 & 01011111111111111111010111 & 0 & 2.4 & 0 & 2.4 & 2.4 & 4 & 4 & 4 & 4 & 76 & 76 & 76 & 76 & 100 & 100 & 100 & 155 & 155 & 155 & 155 & 0 & 152.8 & 0 & 350 & 400 & 400 & 35143.8 \\
17 & 2530 & 01110001111111111111001111 & 0 & 2.4 & 2.4 & 2.4 & 0 & 0 & 0 & 4 & 4 & 76 & 76 & 76 & 76 & 100 & 100 & 100 & 155 & 155 & 155 & 155 & 0 & 0 & 140.8 & 350 & 400 & 400 & 34334.6 \\
18 & 2500 & 11011101111111111111001111 & 2.4 & 2.4 & 0 & 2.4 & 2.4 & 4 & 0 & 4 & 4 & 76 & 76 & 76 & 76 & 100 & 100 & 100 & 155 & 155 & 155 & 155 & 0 & 0 & 104.4 & 350 & 400 & 400 & 33821.5 \\
19 & 2550 & 01011111111111111111010111 & 0 & 2.4 & 0 & 2.4 & 2.4 & 4 & 4 & 4 & 4 & 76 & 76 & 76 & 76 & 100 & 100 & 100 & 155 & 155 & 155 & 155 & 0 & 152.8 & 0 & 350 & 400 & 400 & 35143.8 \\
20 & 2600 & 11001011111111111111001111 & 8.7 & 5.9 & 0 & 0 & 2.4 & 0 & 4 & 4 & 4 & 76 & 76 & 76 & 76 & 100 & 100 & 100 & 155 & 155 & 155 & 155 & 0 & 0 & 197 & 350 & 400 & 400 & 36210.3 \\
21 & 2480 & 11001001111111111111010111 & 2.4 & 2.4 & 0 & 0 & 2.4 & 0 & 0 & 4 & 4 & 76 & 76 & 76 & 76 & 100 & 100 & 100 & 155 & 155 & 155 & 155 & 0 & 90.8 & 0 & 350 & 400 & 400 & 33133.9 \\
22 & 2200 & 11000000000110011111110111 & 2.4 & 2.4 & 0 & 0 & 0 & 0 & 0 & 0 & 0 & 0 & 0 & 76 & 76 & 0 & 0 & 100 & 155 & 155 & 155 & 155 & 96.4 & 76.8 & 0 & 350 & 400 & 400 & 28214.2 \\
23 & 1840 & 01010100000100001111000111 & 0 & 2.4 & 0 & 2.4 & 0 & 4 & 0 & 0 & 0 & 0 & 0 & 61.2 & 0 & 0 & 0 & 0 & 155 & 155 & 155 & 155 & 0 & 0 & 0 & 350 & 400 & 400 & 20464.7 \\
\bottomrule
\end{tabular}
}
\end{table*}


\begin{figure*}
    \centering
    \includegraphics[width=0.95\linewidth]{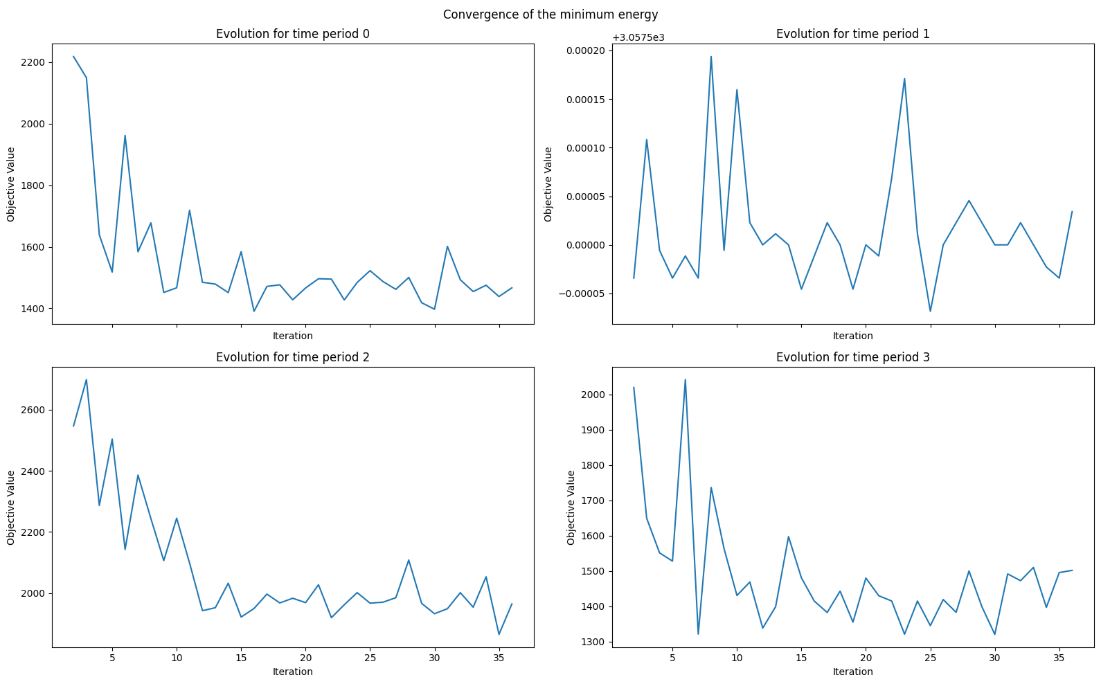}
    \caption{Convergence of our VQA running on IonQ Forte as it approximately solves an hourly problem with $3$ generating units. For this experiment we used our layered ansatz with a single layer and $1,000$ shots per iteration. For this experiment we used a brickwork layout ansatz with $2$ layers of parameterized RY gates and we initialized the parameters by drawing uniformly at random in the interval $[-2\pi, 2\pi]$.}
    \label{fig:3-generator 4-hour Forte}
\end{figure*}

{
\bibliographystyle{IEEEtran}
\bibliography{reference}
}

\end{document}